\documentclass[twocolumn,preprintnumbers,amsmah,amssymb,nopacs,showkeys]{revtex4}
\usepackage{graphicx}
\usepackage{dcolumn}
\usepackage{bm}
 \usepackage[latin1]{inputenc}
\usepackage{multirow}
\usepackage{amsmath}
\usepackage{appendix}

\begin{document}                

\title{Neutron methods for the direct determination of the magnetic induction in thick films}

\author{S.V. Kozhevnikov$^{a, \footnote{kozhevn@nf.jinr.ru}}$, F. Ott$^{b,c}$,  F. Radu$^d$}
\affiliation{$^{a}$Frank Laboratory of Neutron Physics, Joint Institute for Nuclear Research, 141980 Dubna, Russian Federation }
\affiliation{$^{b}$CEA, IRAMIS, Laboratoire L\'eon Brillouin, F-91191 Gif sur Yvette, France}
\affiliation{$^{c}$CNRS, IRAMIS, Laboratoire L\'eon Brillouin, F-91191 Gif sur Yvette, France}
\affiliation{$^{d}$Helmholtz-Zentrum Berlin f\"ur Materialien und
Energie, Albert-Einstein Strasse 15, D-12489, Berlin, Germany}
\date{\today}

\begin{abstract}
We review different neutron methods which allow extracting directly the value of the magnetic induction in thick films: Larmor precession, Zeeman spatial beam-splitting and neutron spin resonance. Resulting parameters obtained by the neutron methods and standard magnetometry technique are presented and compared. The possibilities and specificities of the neutron methods are discussed.
\end{abstract}
\maketitle

\section{Introduction}
Magnetic nanostructures (e. g. thin films, wires, rods, quantum dots) are at the focus of research today because they exhibit novel fundamental properties which are exploited in a broad range of applications. Particularly, the reduced material volume of low-dimensional magnetic systems poses real challenges to neutron based techniques. Therefore, development of new methods for the characterizations of magnetic structures at nanoscale is important, especially as new neutron spallation sources are being built. 
Neutrons have a magnetic moment, isotopic sensitivity and high penetration ability. Therefore neutron scattering is a powerful method for the investigation of magnetic materials, polymers and biological systems and for nondestructive control of materials in bulk. Polarized Neutron Reflectometry (PNR) [1] is successfully used for the investigation of thin magnetic films of thickness smaller than 100 nm and down to 0.1 nm [2]. PNR can extract the absolute magnitude and the direction of the magnetization in each thin layer in a multilayer structure with a high spatial sensitivity of about 0.1 nm in the direction perpendicular to the film surface. However conventional specular reflectometry cannot easily resolve thicknesses which are larger than 100 nm and remains insensitive to the in-plane structures. Nevertheless, the sensitivity to in-plane structures is provided by off-specular neutron scattering [3]. Scattered Intensity appears in the reciprocal space around the specular line when a film has in-plane structures in the direction along the beam propagation, with scales ranging from 600 nm to 60 $\mu$m. Grazing Incidence Small-Angle Neutron Scattering (GISANS) [4,5] can be used when the film has in-plane structures in the direction perpendicular to the beam path. For this case the momentum transfer takes place in a scattering plane which is perpendicular to the incidence plane and provides sensitivity to inhomogeneities in the range 3 to 100 nm. The main peculiarity of the neutron methods mentioned above is that they do not provide direct information on the magnetic and structural properties, but rely largely on model dependent algorithms. This means that we cannot retrieve the structure of the investigated object directly from the neutron intensity. It is the consequence of the fact that the phase of the neutron wave function is lost during the elastic neutron interaction with matter. To describe the measured reflectivity, a model of the investigated system must be provided. Then the calculated reflectivity corresponding to this model is compared with the experimental data. The initial model is further refined until one model describes best a data set. The best model corresponds to the minimum of difference between calculated and experimental data. 
Here we review three different neutron methods which provide a direct measurement of the magnetic induction from experimental data: Larmor precession, Zeeman spatial beam-splitting and neutron spin resonance (NSR).

\begin{figure}[ht]
       \includegraphics[clip=true,keepaspectratio=true,width=1\linewidth]{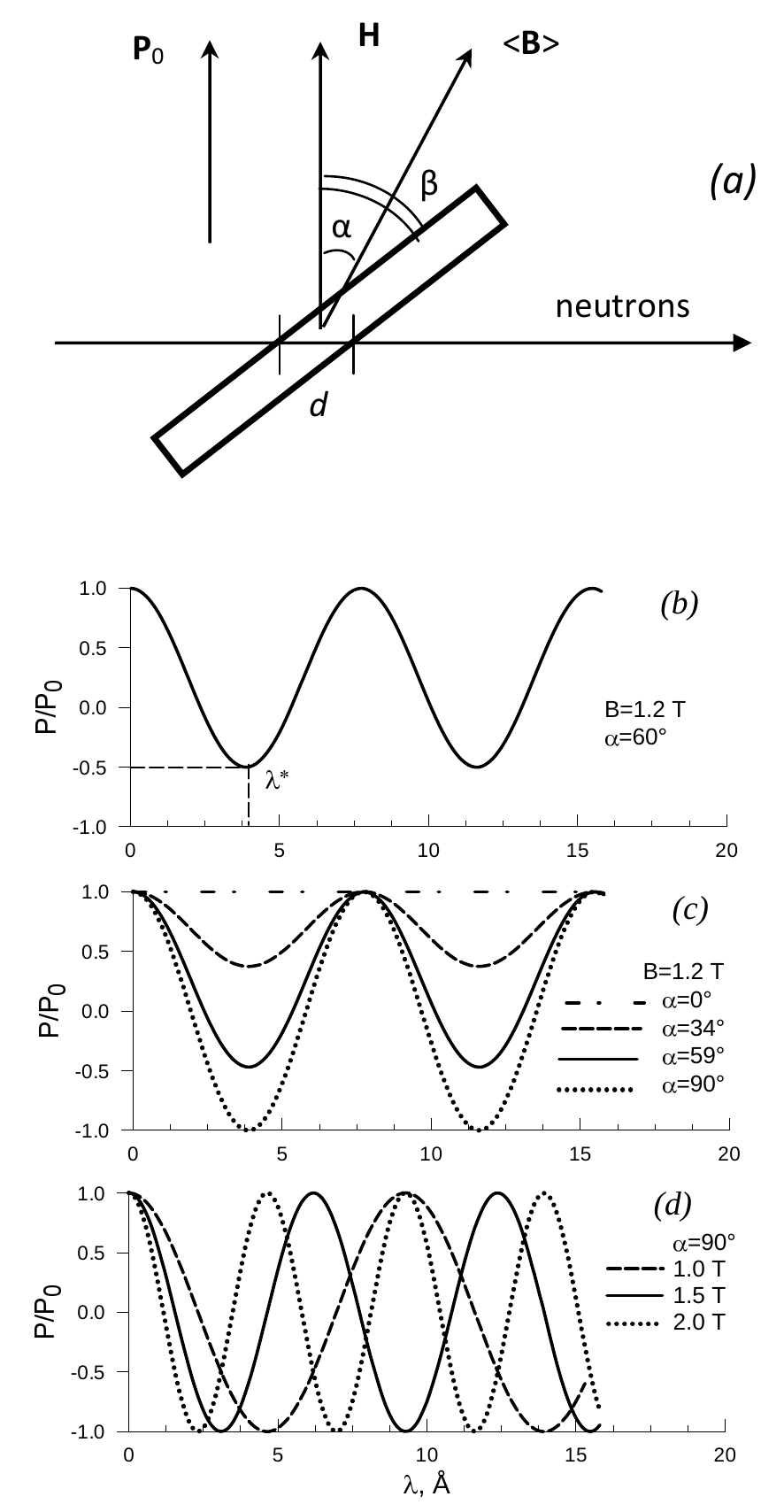}
          \caption{(a) Geometry of the Larmor precession experiment. (b) Calculated normalized polarization as a function of the neutron wavelength for the parameters B=1.2 T, d=14.62 $\mu$m, $\alpha=60^\circ$.  (c) Calculated normalized polarization for the parameters B=1.2 T and d=14.62 $\mu$m and different angles $\alpha$. (d) Calculated normalized polarization for the parameters  $\alpha=90^\circ$ and d=14.62 $\mu$mand different values of B.}
   \label{fig1}
 \end{figure}

\section{Larmor precession method\label{sec2}}

The neutron interaction with a magnetic media contains nuclear and magnetic contributions. To extract a magnetic component which is rather small as compared to a nuclear one, polarized neutrons are used. The polarization degree of the neutron beam is defined as $P=\frac{N^{+}.N^{-}}{N^{+}+N^{-}}$  , where $N^{+}$  and $N^{-}$  are respectively the number of neutrons with spin parallel and antiparallel to the direction of the applied external magnetic field. The polarization of a fully polarized beam is equal to $\pm$1. The polarization of an unpolarized beam is equal to 0. Partially polarized neutron beams have a polarization $0<|P|<1$ . The process of decreasing the polarization degree is termed as \textit{depolarization}.

The theory of neutron depolarization during its passage through the magnetic domains  was developed by Halpern and Holstein [6].  Neutron depolarization was proposed by Rekveldt as a method of investigation of bulk domain structures [7]. This method of depolarization was widely used in 1980-90s for the investigation of ferro- and ferrimagnetics and high-Tc superconductors. Inside a magnetic cluster, Larmor precession of the neutron spin around the vector of local magnetic induction takes place. From the exponential depolarization curve, a dispersion of the magnetic induction may be extracted. This method is however again model-dependent.

The particular case of depolarization at Larmor precession has been proposed as an experimental method using time-of-flight (TOF) technique [8,9]. This method was used for the investigation of magnetic amorphous ribbons [10]. A similar method using fixed wavelength was used for the investigation of very soft permalloy films [11,12].

The scheme of the measuring method is presented in Fig. 1a. A magnetic film is placed under an angle $\beta$ to the external applied magnetic field ${\bf H}$. The angle between the external applied magnetic field and the average magnetic induction $<\bf B>$ in the film is $\alpha$. The neutron path inside the film is d. The initial polarization $\bf P_0$ is directed along the applied field ${\bf H}$ and the angle between the vectors $\bf P_0$ and $<\bf B>$ is $\alpha$. The component of the polarization perpendicular to $<\bf B>$ leads to a Larmor precession of the neutron spin in the mean internal field <B>. The Larmor precession frequency depends on the value of the magnetic induction as $\omega_L=\gamma <B>$ , where $\gamma_n=1.83\times10^{8} s^{-1}T^{-1}$ is the gyromagnetic ratio of the neutron. The angle of Larmor precession in the film is $\varphi=\omega_Lt=\gamma <B>d  / \nu$, where $t$ is the travelling time along the path $d$ inside the film for neutrons of velocity $\nu$. From the expression of momentum, it follows that $\frac{1}{\nu}=\frac{m}{h}\lambda$ , where $m$ is the neutron mass,$h$ is the Plank's constant and $\lambda$ is the neutron wavelength. Thus, the angle of precession in the neutron wavelength scale can be written as $\varphi=\omega_\lambda \lambda$ , where $\omega_\lambda = 0.04633 <B>d$  , with  $\omega_\lambda$  in [\AA$^{-1}$], $<B>$ in [$T$], d in [$\mu$m]. From simple geometric considerations, we obtain the following expression for the case of a uniform magnetic induction $<\bf B>=\bf B$:
\begin{equation}
                     P(\lambda)/P_0(\lambda)=\cos^2{\alpha}+\sin^2{\alpha}\cos{\varphi}=\cos^2{\alpha}+\sin^2{\alpha}\cos{(\omega_\lambda \lambda)}      
\label{eq1}
 \end{equation}
In the case of magnetic fluctuations along the neutron path, additional depolarization coefficients $D_\parallel$ and $D_\perp$ must be introduced in (\ref{eq1}): $ P(\lambda)/P_0(\lambda)=D_\parallel\cos^2{\alpha}+D_\perp\sin^2{\alpha}\cos{\varphi}$ . In the theory of Halpern-Holstein [4] the depolarization coefficients are approximated by exponentials:
 \begin{equation}
                     P(\lambda)/P_0(\lambda)=\exp{(-A_1\lambda^2)}\cos^2{\alpha}+\exp{(-A_2\lambda^2)}\sin^2{\alpha}\cos{(\omega_\lambda \lambda)}      
\label{eq2}
 \end{equation}
where $A_1$ and $A_2$ are constants depending on the mean square induction fluctuations  $<B^2_\parallel>$ and $<B^2_\perp>$ , and the mean size of inhomogeneities. The experimental results of measurements on systems with magnetic inhomogeneities are described in details in [9].

Here we consider the case of a uniformly magnetized film with a magnetic induction $B=<B>$. In Fig. 1b, the normalized polarization is calculated as a function of the neutron wavelength. The parameters for the calculations are the following: B=1.2 T, d=14.62 $\mu$m, $\alpha=60^\circ$. One can see that the first minimum of the periodic oscillation dependence occurs at the characteristic wavelength $\lambda^*$. This point corresponds to a precession of  $\varphi=\omega_\lambda\lambda^*=\pi$ . From this condition we can define the value of the magnetic induction: 
 \begin{equation}
                    B=\frac{\pi}{0.04633 d\lambda^*}   
\label{eq3}
 \end{equation}

  From the equation (\ref{eq1}) we can extract the angle $\alpha$ as:
 \begin{equation}
                    \cos{\alpha}=\sqrt{\frac{1+  P(\lambda^*)/P_0(\lambda^*)}{2}}    
\label{eq4}
 \end{equation}

In Fig. 1c, the normalized polarization is calculated for different angles $\alpha$ at $B=1.2$~T and $d=14.62~\mu m$ as a function of the neutron wavelength. One can see that at $\alpha=90^\circ$ (for which the vector of magnetic induction is perpendicular to the applied magnetic field) the amplitude of the polarization at the minimum at $\lambda^*$ is maximal. When decreasing $\alpha$ angles, the amplitude is decreased and becomes zero at $\alpha=0^\circ$ (there is no Larmor precession when the magnetic induction is parallel to the applied field).

In Fig. 1d, the normalized polarization is calculated for different values of the magnetic induction $B$ at $\alpha=90^\circ$ and $d=14.62~\mu m$ as a function of the neutron wavelength. One can see that the period of the Larmor precession is decreasing as $B$ increases.    
      
Let us consider experimental results. A $5 \mu m$ thick magnetic film Fe(86 at\%)Al(9.6\%)Si(4.4\%) on a ceramic substrate which is 10(along the beam)x20x1$ mm^3$ was investigated at the polarized neutron spectrometer SPN-1 (reactor IBR-2, JINR, Dubna, Russia). IBR-2 is a pulsed reactor and therefore the TOF technique was used for these measurements. An external magnetic field ${\it H}=4.5$~kOe was applied under an angle $\beta=70^\circ$ with respect to the sample surface (see Fig.1a). The angle between the neutron beam and the sample surface was $20^\circ$. Thus, the path of the beam inside the film was  $d=5/\sin20^\circ=13.62~\mu$m. The experimental setup was the following: a neutron polarizer, a spin-flipper of Korneev type, the sample was placed in the electromagnet, an analyzer and a $^3$He gas monodetector [13,14]. The polarizer and the analyzer were made of 5 m long neutron guides, each consisting of two curved magnetic mirrors [14,15]. At the exit of the polarizer, the polarized neutron beam is shaped by a narrow vertical slit of 2.5x100~$mm^2$. The sample was placed also vertically and masked with a diaphragm of 2.0x15~$mm^2$. The distances were the following: the distance polarizer-sample was 3~m, the distance sample-analyzer was 2.27~m, the distance sample-detector was 8~m, the TOF base was 37~m.

The experimental curves of the normalized polarization as a function of the neutron wavelength at different external fields are presented in Fig. 2a. The points are the experimental data and lines are the calculated curves for the experimentally obtained parameters. One can see a minimum at $\lambda^*$ corresponding to half the period of Larmor precession. This minimum moves to the left side as the value of the applied field is increasing. From this experimental data we extracted directly the value of the magnetic induction (Fig. 2b) and the angle $\alpha$ (Fig. 2c) as a function of the applied field value. In low applied fields of 70~Oe, the magnetic induction vector is directed parallel to the film plane due to demagnetizing factor ($\alpha = \beta = 70^\circ$). The point at 4.5~kOe corresponds to the data for FeAlSi film of the thickness of 20~$\mu$m extracted from the spatial beam-splitting described in section \ref{sec3}. One can see that the beam-splitting data overlap with the Larmor precession data. The method of Larmor precession described above can be used for the direct determination of magnetic induction values. 
In the next section we will present the method of Zeeman spatial beam-splitting which takes place at an interface in magnetically non-collinear systems. 

\begin{figure}[ht]
       \includegraphics[clip=true,keepaspectratio=true,width=1\linewidth]{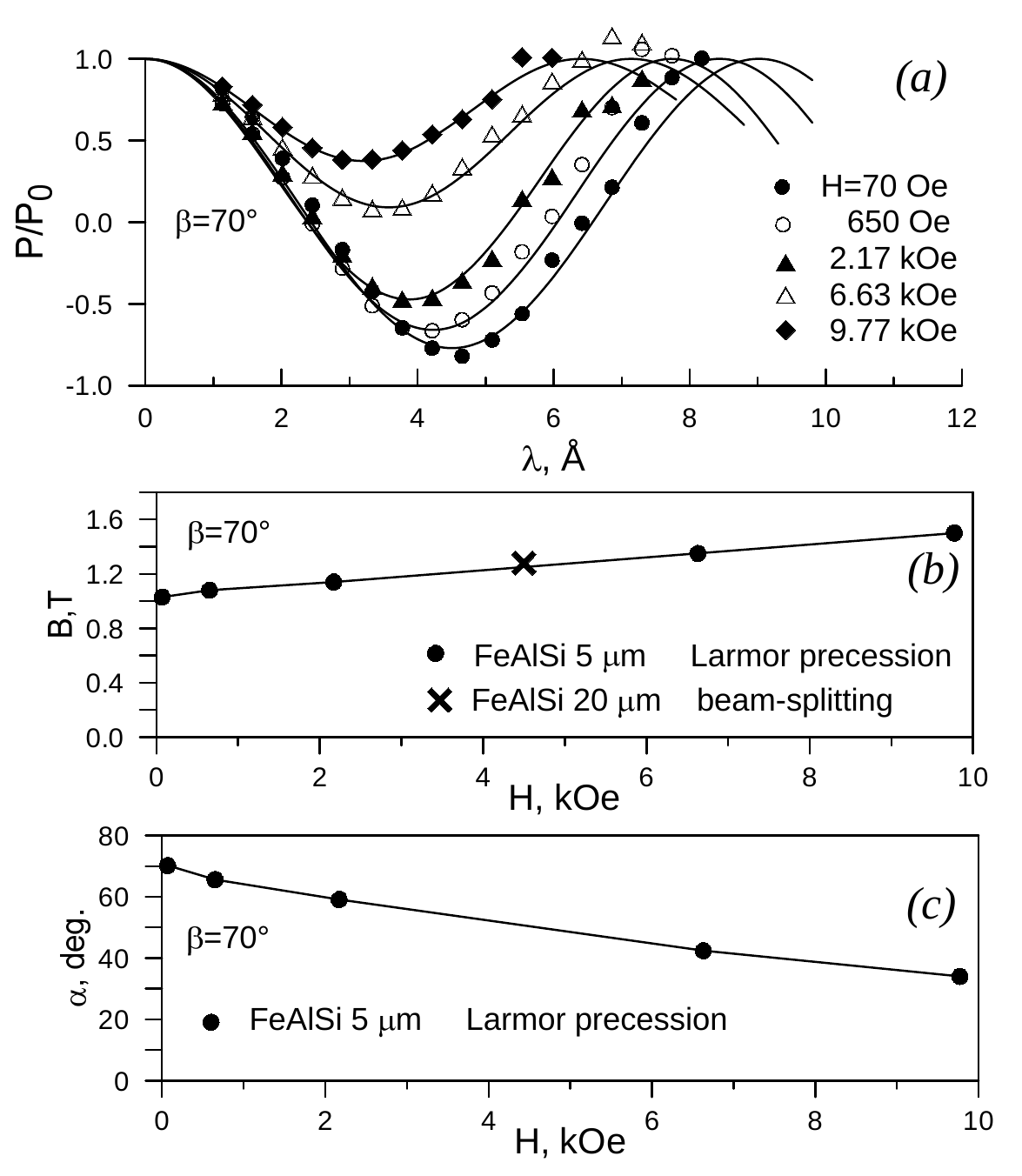}
          \caption{Experimental results by Larmor precession method. (a) Measured polarization as a function of the neutron wavelength. (b) Extracted experimental value of the magnetic induction as a function of the applied magnetic field. (c) Defined experimental value of the angle between the vectors of applied field and magnetic induction in the film. }
   \label{fig2}
 \end{figure}

\section{Zeeman spatial beam-splitting \label{sec3}}

Zeeman spatial beam-splitting takes place only in magnetically non-collinear systems in which there is a sharp change of orientation between the external applied field (quantization field) and the magnetic induction in the magnetic system. Several such experimental situations are presented in Fig. 3. Magnetic thin films exhibit large shape anisotropy, forcing the magnetization to be oriented in the film plane. Thus it is possible to apply rather large external fields $\bf B_0$ perpendicular to the film while the magnetization essentially remains in the film plane (Fig. 3a). When the film has very large in-plane anisotropy the external magnetic field $\bf B_0$ may be applied in the film plane but perpendicular to the anisotropy axis so that the magnetization and $\bf B_0$ are not collinear (Fig. 3b). In a ripple domain structure (Fig. 3c) there is an angle between the applied magnetic field $\bf B_0$ and the induction $\bf B$ in the domains. In systems with a perpendicular anisotropy (Fig. 3d), the external magnetic field $\bf B_0$ can be applied in-plane while the vector of magnetization $\bf M$ remains essentially perpendicular to the sample surface.

Spin-flip and beam-splitting at the boundary of two magnetically non-collinear media were predicted theoretically [16] and observed experimentally in a reflection geometry [17,18]. The theory of spin-flip was developed in [19]. Beam-splitting at refraction was observed in [20]. The review on experiments with beam-splitting was done in [21,22]. The Zeeman beam-splitting in domain and cluster structures was investigated in [23-25].

\begin{figure}[ht]
       \includegraphics[clip=true,keepaspectratio=true,width=1\linewidth]{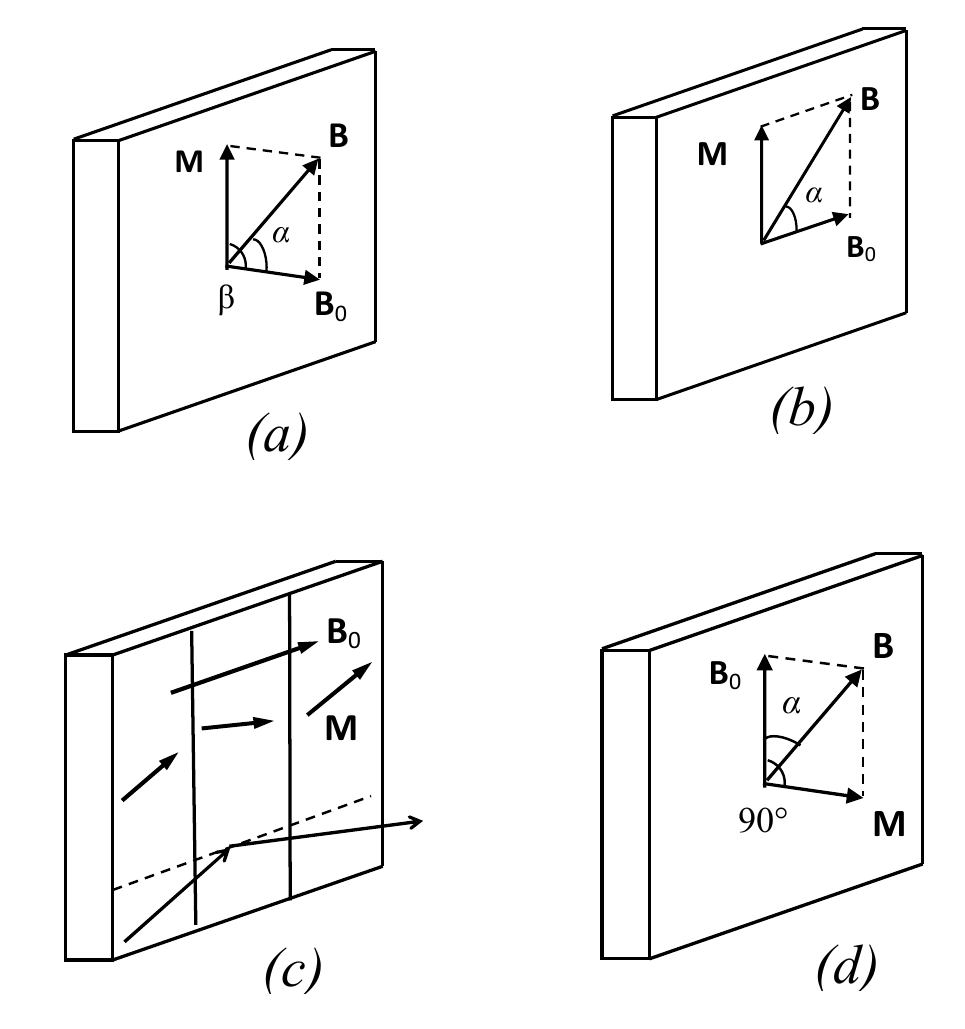}
          \caption{Magnetically non-collinear systems: (a) shape anisotropy; (b) in-plane anisotropy; (c) domain structure; (d) perpendicular anisotropy.
}
   \label{fig3}
 \end{figure}

    Let us consider shortly the principle of the Zeeman spatial beam-splitting. A detailed description can be found in [22]. A magnetic film with in-plane magnetization is placed in an external magnetic field $\bf B_0$ which is applied under an angle $\beta$ with respect to the sample surface (Fig. 3a). The vector of the magnetic induction in the film is $\bf B$. The neutron beam enters to the sample surface under a glancing angle $\theta_i$  (Fig. 4a). The glancing angle of the reflected beam is  $\theta_f$. During the reflection, the neutrons experience a spin-flip which takes place with some probability depending on the angle $\alpha$ between vectors $\bf B_0$ and $\bf B$ as $W\sim \sin^2{\alpha}$    (Fig. 4a). 

\begin{figure}[ht]
       \includegraphics[clip=true,keepaspectratio=true,width=1\linewidth]{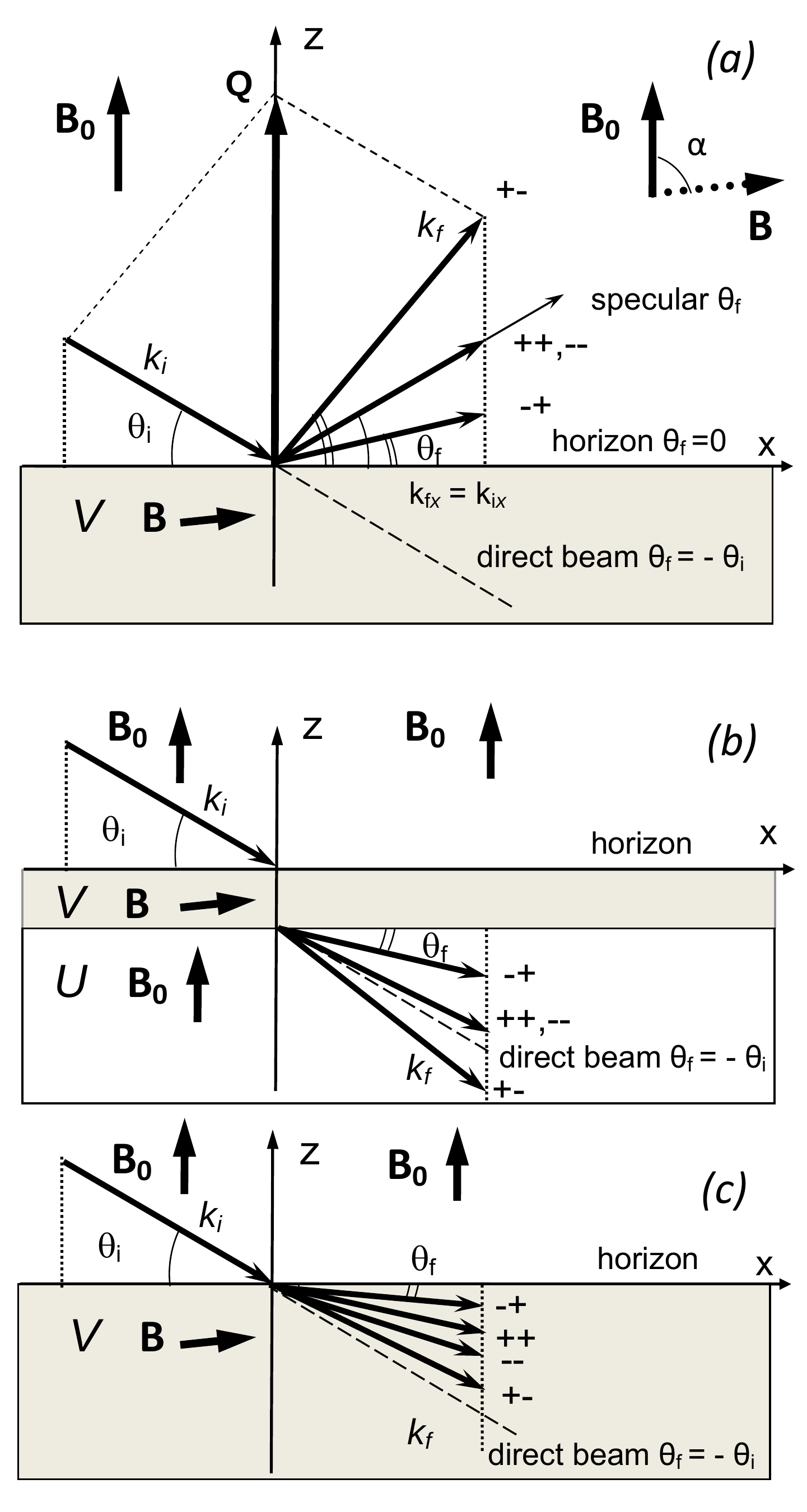}
          \caption{Scheme of the spatial beam-splitting in a homogeneous magnetic film when the in-plane component of neutron wave vector is conserved: (a) reflection from the film; (b) transmission through a thin magnetic film and refraction in a nonmagnetic substrate; (c) refraction at a boundary of a thick magnetic film.}
   \label{fig4}
 \end{figure}

The energy conservation for a spin-flip process from the initial spin '+' to the final spin '-' allows writing:

 \begin{equation}
                    \frac{\hbar k_i^2}{2m} +\mu B_0=  \frac{\hbar k_f^2}{2m}-\mu B_0
\label{eq5}
 \end{equation}
where $k_i$ and $k_f$ are the incident and reflected neutron wave-vectors, $\mu$ is the neutron magnetic moment and $m$ is the neutron mass. The neutron experiences an energy change equal to $2\mu B_0$ which modifies its outgoing wave-vector $k_f$. Because of the continuity of the in-plane wave-vector component  $k_{ix}=k_{fx}$, the total energy is transferred on the perpendicular component $k_z$ (see Fig. 4a). This change in the scattering wave-vector leads to a measurable change in the reflection direction off the specular direction which is easily observable in neutron reflectivity experiments. Note that the neutron wavelength is also changed ($\Delta k/k=\Delta\lambda/\lambda\approx 10^{-3}$ ) but this change is not measurable on conventional instruments. We also stress that this case of  `off-specular` scattering has a very different origin from the diffuse scattering due to in-plane sample microstructures for which the change in the scattering wave-vector $\bf Q$ is due to an in-plane momentum transfer. We introduce the notations $p_{i,f}=k_{i,f} \sin\theta_{i,f}$  in order to simplify the mathematical expressions:
\begin{equation}
                    \frac{\hbar p_i^2}{2m} +\mu B_0=  \frac{\hbar {(p^{+-}_f)}^2}{2m}-\mu B_0
\label{eq6}
 \end{equation}

As a result, an unpolarized beam is split into three beams at reflection (Fig. 4a):
\begin{equation}
                  ( \theta^{+-}_{f})^2=\theta_{i}^2+2\mu B_0\lambda^2\frac{2m}{h^2}
\label{eq7}
 \end{equation}
\begin{equation}
                   \theta_{f}^{++}= \theta_{f}^{--}= \theta_{i}
\label{eq8}
 \end{equation}
\begin{equation}
                  ( \theta^{-+}_{f})^2=\theta_{i}^2-2\mu B_0\lambda^2\frac{2m}{h^2}
\label{eq9}
 \end{equation}

When the neutron beam is transmitted through a thin magnetic film and refracted in a thick non-magnetic substrate with the nuclear optical potential $U$ (Fig. 4b), the energy conservation law gives the following expression for the '+-' spin-flip process:
\begin{equation}
                   \frac{\hbar p_i^2}{2m} +\mu B_0=  \frac{\hbar {(p^{+-}_f)}^2}{2m}-\mu B_0+U
\label{eq10}
 \end{equation}
In this case, the unpolarized beam is spatially split also into three beams.

At refraction in a thick magnetic film of nuclear potential $V$, the unpolarized neutron beam is split into four beams (Fig. 4c). The energy conservation law is written in Eq. (A.1). Using the Eqs. (A.2-A.5), this type of beam-splitting at refraction on one boundary allows to extract experimentally three parameters: the external magnetic field $ B_0$, the magnetic induction in the film $B$ and the nuclear potential $V$.

In this section we present experimental results on a thick FeAlSi film Fe(86 at\%)Al(9.6 at\%)Si(4.4 at\%). The structure of the sample was FeAlSi(20 $\mu$ m)/Cr(500 \AA)//CaTiO3(substrate). The size of the sample was 5x20x1~mm$^3$ (5 mm along the beam). The external magnetic field 4.5 kOe was applied under an angle $\beta=70^\circ$ with respect to the sample surface (as in Fig. 3a). The incident angle was  . The incident angular divergence was $0.015^\circ$. The measurements were carried out at the polarized neutron spectrometer SPN-2 (reactor IBR-2, JINR, Dubna Russia) which is the modernized version of the spectrometer SPN-1. The experimental setup consisted of 5 m long polarizing neutron guide [14,15], the first spin-flipper of Korneev type, the second radiofrequency adiabatic spin-flipper with large working area of 100 mm in diameter, the curved multislit supermirror analyzer with 40x40~mm$^2$ window and one-dimensional position sensitive 3He gas detector [13,14] with a spatial resolution of 1.5 mm and a window of 40x120~mm$^2$. The sample plane was aligned vertically. The scattering plane was horizontal and the analyzer mirrors were also horizontal. The distances were following: the distance sample-detector was 8 m and the TOF base was 37 m. The polarizing efficiency of the long polarizer was dropping from 98~\% at $\lambda = 2$~\AA to 35 \% at 6~\AA. That is why on the spectra the incident polarized beam consists of a significant part of (parasitic) opposite polarization. The multi-slit analyzer has high polarizing efficiency of 98 \% at $\lambda = 2$~\AA and 95 \% at 6~\AA. The parameters of some parts of this setup can be found in [14]. The spin-flippers modes UU, UD, DU and DD correspond to the intensities '++', '+-', '-+' and '--', respectively. 

In Fig. 5a, the geometry of the experiment is presented. The polarized neutron beam enters through the surface, is refracted on the boundary 'vacuum - film' and goes out from the edge of the thick film FeAlSi. This trajectory is marked by the index 1. The solid and dashed lines correspond to the initial spin '+' and '-', respectively. In Fig. 5b, the calculated trajectories ($\lambda, \theta_f$)  are presented. The symbols correspond to the experimental data. The experimental parameters $B_1$, $B_0$ and $V_1$ were extracted from the experimental data using Eqs. (A.2-A.5). One can see four refracted beams '-+', '++', '--' and '+-' between the horizon and the direct beam direction. 

The index 2 marks the beam which enters through the edge of the thick film, is further refracted and goes out from the edge of the non-magnetic substrate. The thin nonmagnetic Cr (500 \AA) film does not disturb the refracted beam. The solid and dashed lines correspond to the initial spin '+' and '-' states, respectively. The curves in Fig. 5b for this trajectory were calculated for the experimental parameters $B_2$, $B_0$ and ($V_2-U$) obtained from Eqs. (B.2-B.5). One can see three refracted beams below the direct beam direction. The beam '-+' for this trajectory 2 is weakly refracted and close to the direct beam direction.

In Figs. 5c-f, the neutron intensity of the refraction 1 at the interface 'air-film' is presented as a function of a final glancing angle for different states of two spin-flippers UU, DD, DU and UD. The intensity was integrated over the neutron wavelength interval from 2.5 to 3.0~\AA. In Fig. 5c, for the UU state, one can see the large peak corresponding to the direct beam and the '++' refracted beam peak marked by an arrow. In Fig, 5d, for the DD state, one can see the high peak of the direct beam and the '--' refracted beam peak marked by the arrow. For the spin-flippers state DU, the neutron intensity is shown in Fig. 5e. The weak peaks '--', '++' and '-+' are marked by arrows. The main peak is '-+'. The peaks '--' and '++' correspond to parasitic neutrons due to the imperfection of the polarizer and the analyzer. In the region of the direct beam in Fig. 5e, the neutron intensity is much higher than for the spin-flippers state UD in Fig. 5f. It means that the beam '-+' of the refraction 2 on the interface 'film-substrate' is overlapped by the direct beam as it is calculated in Fig. 5b. For the spin UD in Fig. 5f, one can see the weak trace of the direct beam peak, the main peak '+-' and the weak trace of the intensity for '--' and '++' spin states. Figures 5c-f demonstrate that an unpolarized neutron beam is split into four beams at refraction on one boundary of two magnetically non-collinear media [26].

\begin{figure}[ht]
       \includegraphics[clip=true,keepaspectratio=true,width=1\linewidth]{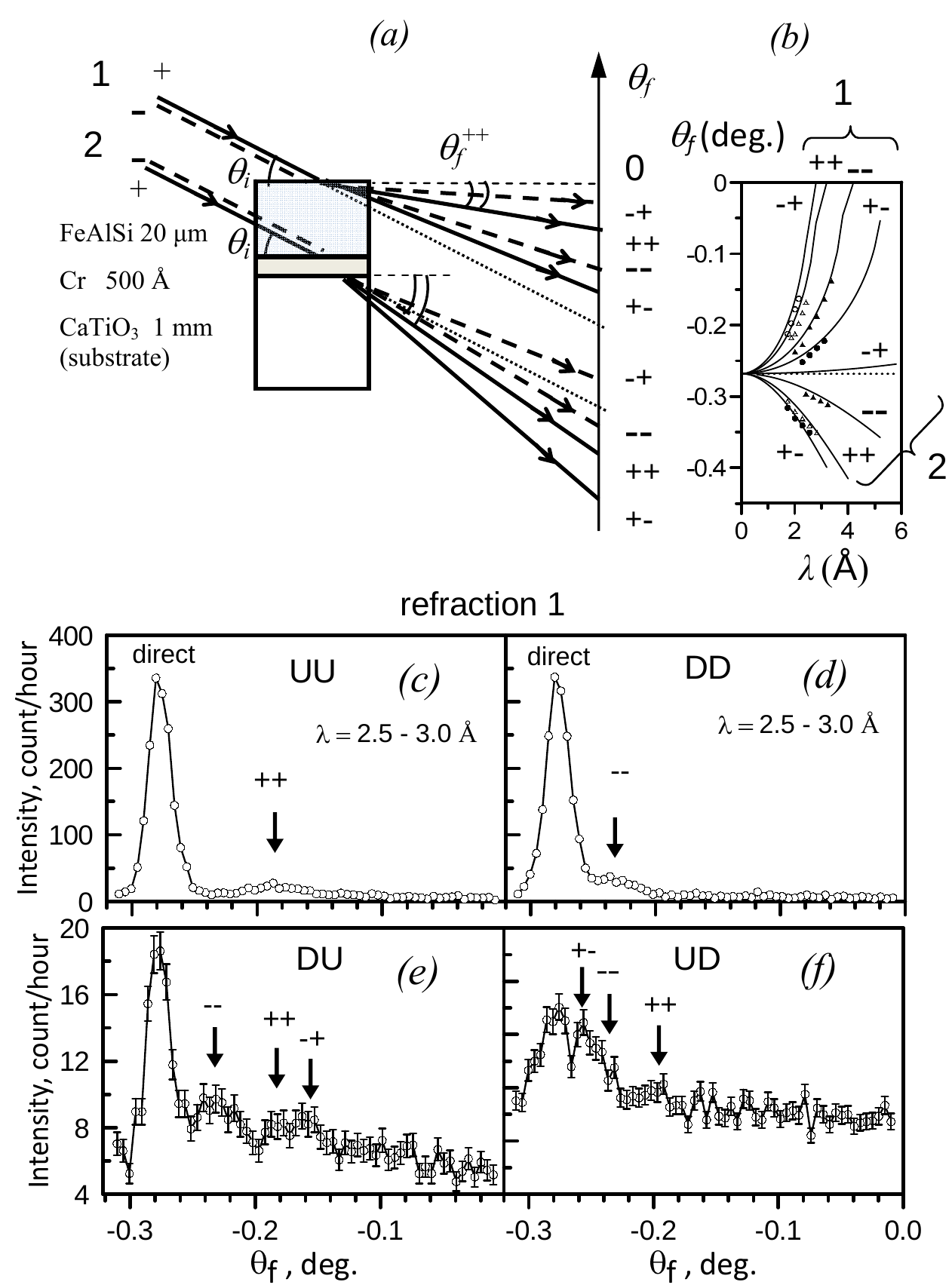}
          \caption{ (a) The geometry of the refraction experiment. Index 1 marks the refraction on the boundary 'air-film' and index 2 corresponds to the refraction 'film-substrate'. (b)  Trajectories of the angle of refracted beams 1 and 2 as a function of the neutron wavelength (calculation is solid line and experiment is symbols). (c)-(f) The neutron intensity of refracted beams on the boundary 'air-film' as a function of a final glancing angle for the different spin-flippers states UU, DD, DU and UD. The intensity is integrated in the neutron wavelength interval 2.5 - 3.0 \AA. }
   \label{fig5}
 \end{figure}

The experimental two-dimensional map  ($\lambda, \theta_f$)   is presented in Fig. 6 for the spin-flippers states UU (a), DU (b), DD (c) and UD (d). The refracted beams are indicated by arrows and the indices 1 for the refraction 'air-film' and 2 for the refraction 'film-substrate'. One can see that the angular beam-splitting increases as the neutron wavelength increases. The same data in the coordinates ($_p12_i-p_f^2, p_i^2+p_f^2$)  are shown in Fig. 6e-h. This presentation is more suitable for the extraction of the parameters of the investigated system because of the refracted beams follow horizontal lines. The experimental data in Fig. 6e-h are presented in Table~\ref{table1}.

\begin{figure}[ht]
       \includegraphics[clip=true,keepaspectratio=true,width=1\linewidth]{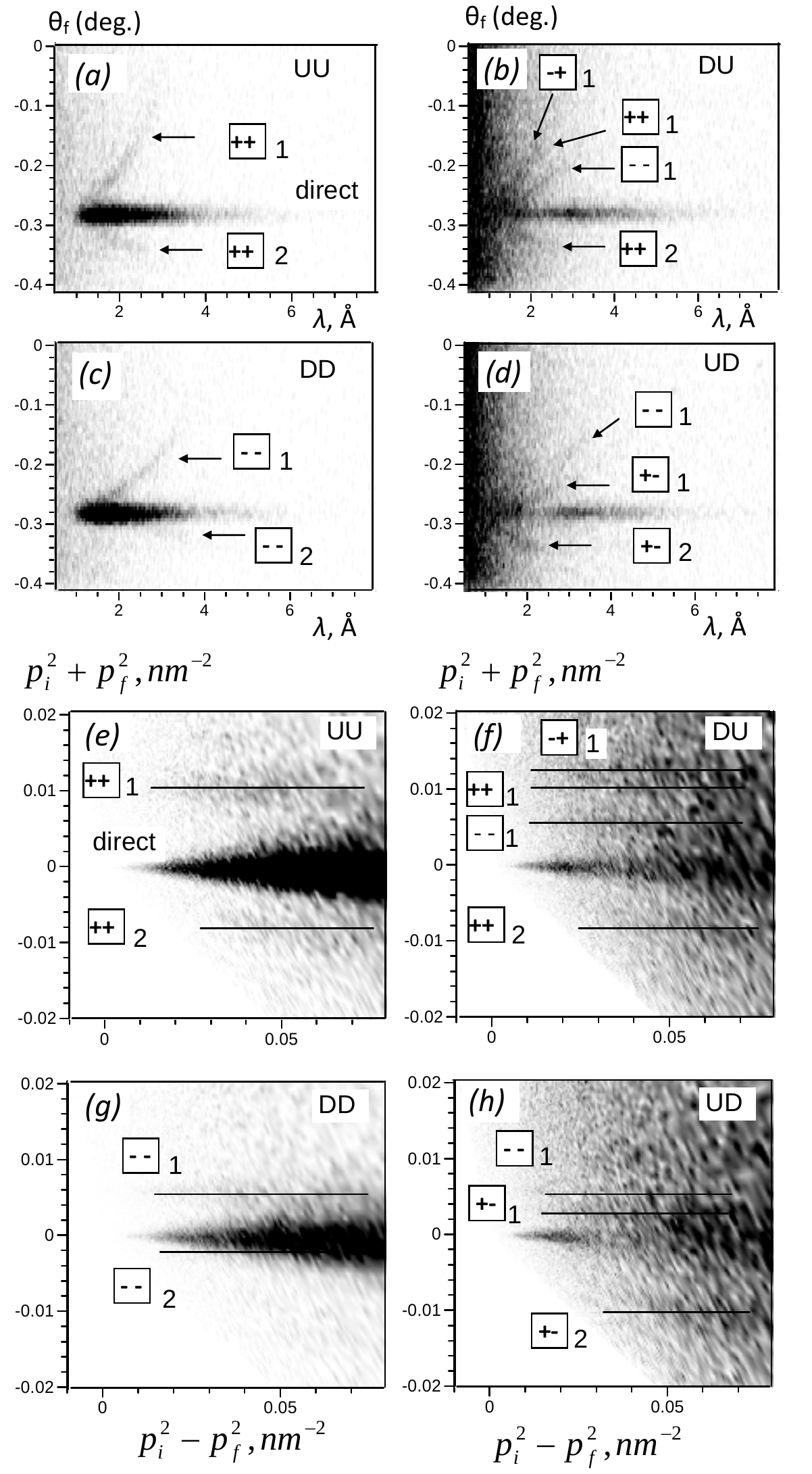}
          \caption{(a)-(d) Experimental two-dimensional map of intensity   for the different spin-flippers states UU, DD, DU and UD. (e)-(h) Experimental two-dimensional map of intensity in the coordinates $(p_i^2-p_f^2,p_i^2+p_f^2)$  . }
   \label{fig6}
 \end{figure}

\begin{table}[h!]
  \begin{center}
    \caption{Experimental data for refraction 1 and 2. }
    \label{table1}
\begin{tabular}{ |c|c|c| } 
\hline
\multirow{2}{*}{spin states}& \multicolumn{2}{c|}{ $p_i^2+p_f^2,[mm^{-2}]$ } \\\cline{2-3}
 &1 & 2 \\ \hline
-+&0.0131&\\ \hline
++&0.0150&-0.0076\\ \hline
--&0.0058&-0.0025\\ \hline
+-&0.0031&-0.0103\\ \hline
  \end{tabular}
  \end{center}
\end{table}

Using the experimental data from the table 1 and eqs. (A.2-A.4), we extracted the parameters of the system for the refraction 1. For example, for the beam '+-' we calculate:
\begin{multline}
B_0=172.3~\{|p_i^2+(p_f^{--})^2|-|p_i^2+(p_f^{+-})^2|\}=0.47~T \\ 
B_1=172.3~\{|p_i^2+(p_f^{++})^2|-|p_i^2+(p_f^{+-})^2|\}=1.28~T
  \notag
\end{multline}
In the same way, we can use the beam '-+'. 
For the nuclear potential of the magnetic film:
$$
V_1=1.039\times10^{3}\{|p_i^2+(p_f^{++})^2|+|p_i^2+(p_f^{--})^2|\}=169~neV
$$
The angular resolution of the incident beam is $\Delta \theta /\theta=5 \%$. We estimate the accuracy of the extracted parameters as 10 \%.  The values of the experimental parameters extracted from the refraction 1 are presented in the Table~\ref{table2}. The tabulated calculated nuclear potential of the FeAlSi film is 187 neV.  One can see that the magnetic parameters $B_0$ and $B_1$ coincide for the beams '+-' and '-+' and nuclear potential $V_1$ is close to the calculated value inside the error bars.

\begin{table}[h!]
  \begin{center}
    \caption{Extracted values for refraction 1 '-+' and '+-'. }
    \label{table2}
\begin{tabular}{ |c|c|c| } 
\hline
parameter&+-&-+\\ \hline
$B_0$,T &0.47(5) & 0.45(5) \\ \hline
$B_1$,T&1.28(13)&1.26(13)\\ \hline
$V_1$,neV& \multicolumn{2}{c|}{ 169(17) }\\ \hline
  \end{tabular}
  \end{center}
\end{table}
 
The parameters extracted from the refraction 2 using eqs. (B.2-B.4) are presented in Table~\ref{table3}.  
One can see that parameter $B_0$ is close to the applied external magnetic field. The value of the magnetic induction is close the value for the refraction 1. It means that the magnitude of the magnetic induction is the same at the two interfaces of the FeAlSi film, i.e. this film is magnetically homogeneous. If we add the tabulated calculated nuclear potential of the CaTiO$_3$ substrate 86 neV to the extracted experimental value ($V_2-U$) we obtain the value of the FeAlSi film nuclear potential $V_2=190$~neV. And this value is close to the tabulated potential 187~neV.

\begin{table}[h!]
  \begin{center}
    \caption{Extracted values for refraction 2. }
    \label{table3}
\begin{tabular}{ |c|c|c| } 
\hline
parameter&value\\ \hline
$B_0$,T &0.47(5) \\ \hline
$B_1$,T&1.34(13)\\ \hline
$V_2-U$,neV&104(10)\\ \hline
  \end{tabular}
  \end{center}
\end{table}

In the section~\ref{sec2}, we presented the parameters of the similar film FeAlSi (but $5~\mu$m thickness) extracted from the Larmor precession method. The value of magnetic induction (see Fig. 3b) for an applied field 4.5 kOe is equal to 1.23 T. Thus, the methods of the Larmor precession and the Zeeman spatial beam-splitting give the same value of the magnetic induction for the same material FeAlSi of the thick films.

In these two sections we have considered uniformly magnetized saturated films. In the next section, we will present a method for the investigation of a domain structure in a non-saturated magnetic film placed in a low external oscillating field.

\section{Neutron spin resonance\label{sec4}}

Neutron spin resonance takes place during the passage of neutrons through an oscillating magnetic field. In this system, the neutron spin flip probability has a resonant behaviour when the Larmor precession frequency of the neutron spin in a constant magnetic field coincides with the frequency of the rotating one. Neutron spin resonance in oscillating magnetic fields was considered theoretically in [27,28]. This phenomenon was used for the experimental measurement of neutron magnetic moment [29] and was applied to fabrication of resonant spin-flippers [30], which have been successfully used in neutron experiments till nowadays.

Neutron spin resonance in matter was considered theoretically in [31,32] and observed experimentally in [33]. The theoretical investigations of the magnetic layered structures placed in crossed constant and oscillating magnetic fields were reported in [34-36]. 

The experiment was conducted at the polarized neutron reflectometer NREX (Forschungs-neutronenquelle Heinz Maier-Leibnitz, FRM II, Garching near Munich, Germany) with a sample plane oriented horizontally. The sample was a 500 nm thick permalloy film deposited on a Si substrate which was 25x25x1~mm$^3$. Permalloy (Py) stands for the magnetic alloy Fe(20.6 at. \%)Ni(79.4 at. \%). A supermirror polarizer in transmission mode provided a polarized incident beam of 97 \%. A Mezei type spin-flipper with 100 \% efficiency was used. A $^3$He two-dimensional position-sensitive detector with a 3 mm spatial resolution was used. The neutron wavelength was 4.26~\AA (1 \% FWHM resolution). The angular resolution of the incident beam was 0.006°. The distance between the first collimating slit and the sample was 2200 mm. The distance 'sample-detector' was 2500 mm.

A radio-frequency generator (Rohde\&Schwarz SMC100A) and a power amplifier (100 W CW, American Microwave Technology M3200/M3000) were used to create an oscillating field in a resonant LC circuit. The adjustable capacity could be varied in the range 5-100 pF. The magnetic coil had a length of 50 mm, a cross-section of 5x30~mm$^2$, and consisted of 50 turns of 1 mm diameter Cu wire.

The geometry of the experiment is shown in Fig. 7a,b. The polarized neutron beam enters the sample surface under a fixed incidence angle $\alpha_i=0.4^\circ$ and is reflected at an angle $\alpha_f$. The sample was placed inside a small radiofrequency coil which produced the oscillating magnetic field $H_1(t)=H_1\cos(\omega t)$  directed parallel to the beam path. The amplitude of the oscillating field was equal to $H_1$=10 Oe and its frequency could be swept in the interval 28-30 MHz. A constant magnetic field $H_0$=20 Oe was produced by an electromagnet and was directed perpendicular to the scattering plane. The spin of the incident beam was parallel (up or '+') or antiparallel (down or '-') to the constant magnetic field.   
  
\begin{figure}[ht]
       \includegraphics[clip=true,keepaspectratio=true,width=1\linewidth]{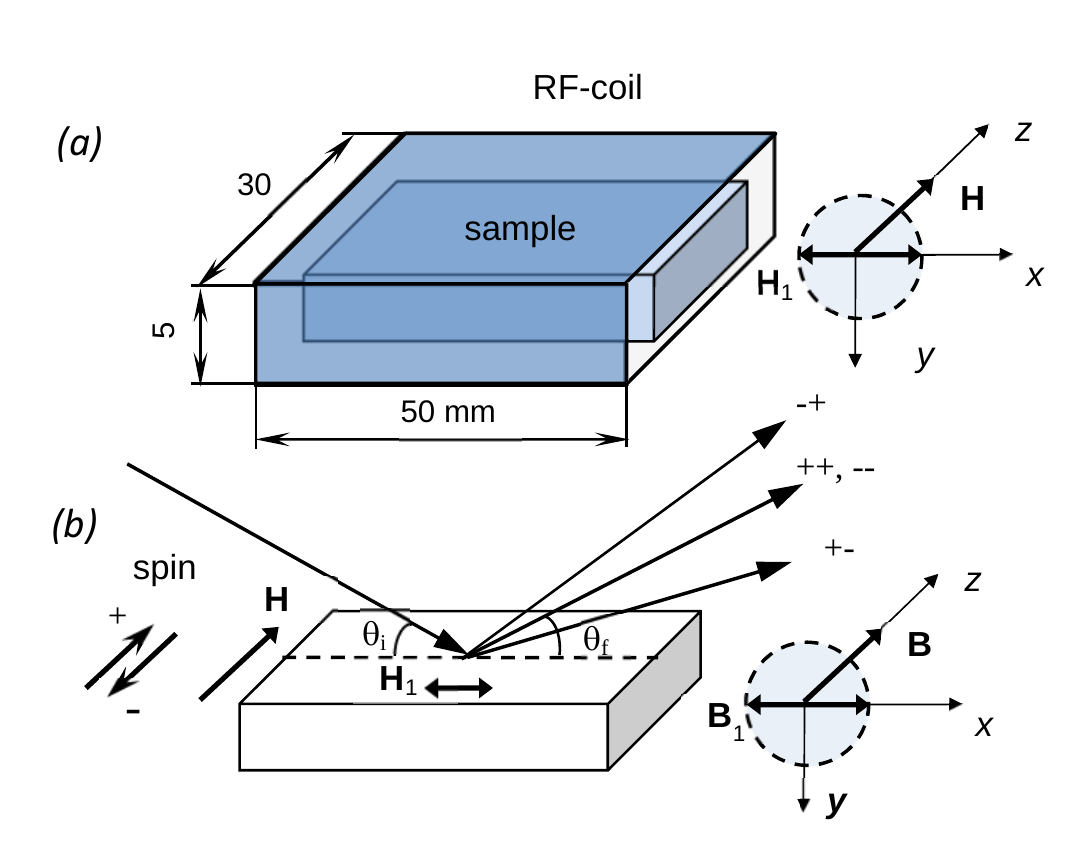}
          \caption{Experimental scheme of neutron spin resonance measurements. (a) Arrangement of the radiofrequency coil. (b) Geometry of the reflection of the beam from the sample. }
   \label{fig7}
 \end{figure}

\begin{figure}[ht]
       \includegraphics[clip=true,keepaspectratio=true,width=1\linewidth]{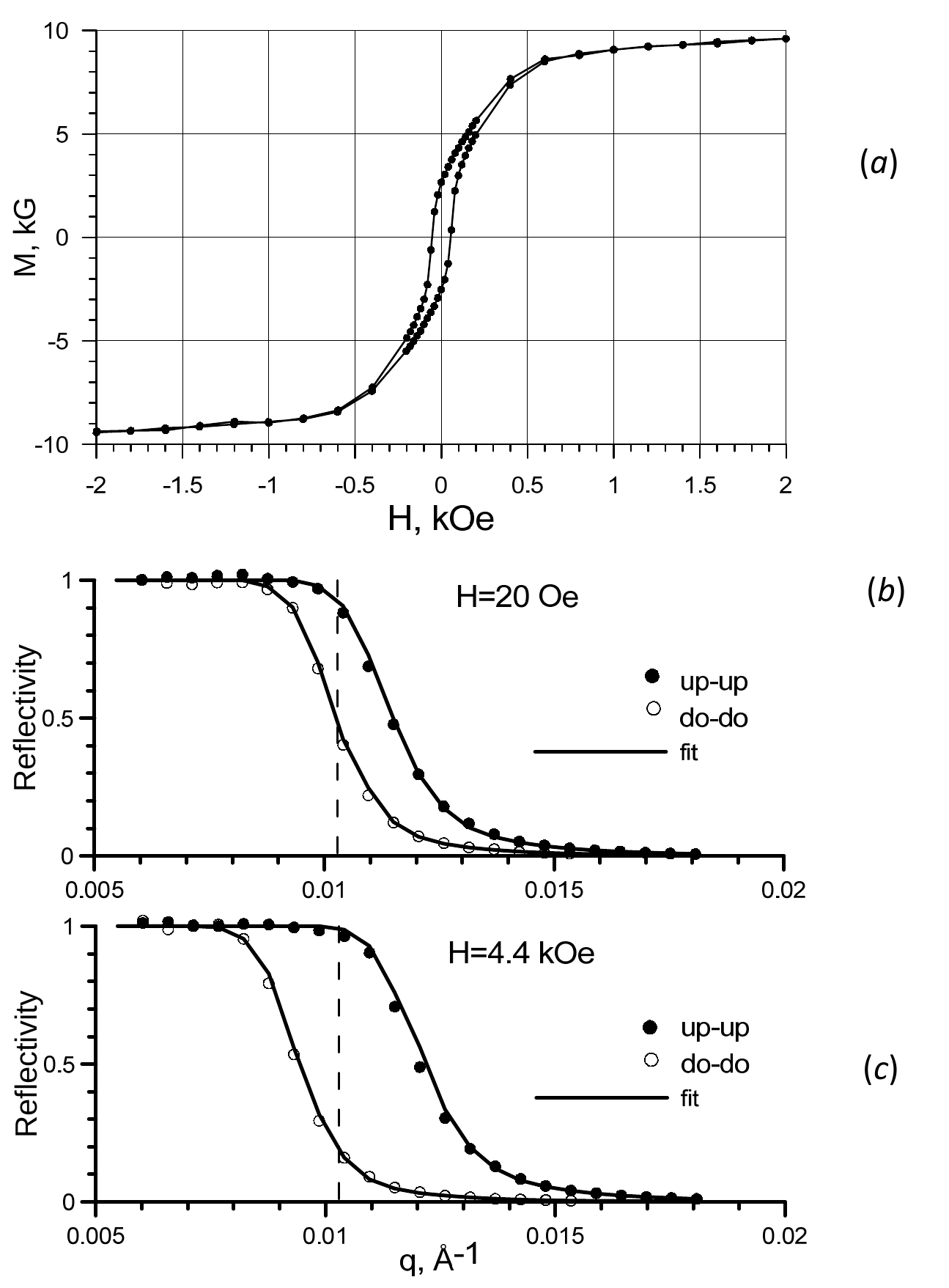}
          \caption{Characterization of the permalloy (500 nm) film: (a) VSM data; (b) neutron reflectivity at 20 Oe applied field; (c) neutron reflectivity at 4.4 kOe applied field.}
   \label{fig8}
 \end{figure}
The hysteresis loop of the sample measured by a Vibration Sample Magnetometer (VSM) is presented in Fig. 8a. The film was saturated at 2 kOe and the magnetization was equal to M=9.5 kG. The reflectivity curves in Fig. 8b,c were measured on a polarized neutron reflectometer PRISM (reactor ORPHEE, LLB, Saclay, France). The neutron wavelength was 4.0~\AA (0.37~\AA FMHW) and the incident angularresolution was 0.04$^\circ$. The fit (lines) was done by the program SimulReflec [37]. The fit result $U = 211.2$~neV for the nuclear potential corresponds to 89 \% of the theoretical value $U = 237.3$~neV for the nominal sample stoichiometry Fe(20.6 \% at.)Ni(79.4 \% at.). In a low external field 20 Oe the film was partially magnetized and the fitted magnetization was of M=4.0 kG (Fig. 8b). In a high external field of 4.4 kOe (Fig. 8c) the film was saturated and had a magnetization value equal to M=9.0 kG. The dashed line corresponds to $q=2 \pi \sin(\alpha_i)/\lambda= 0.01025$~(\AA$^{-1}$) as at the measurements on NREX reflectometer. 
 
The specularly reflected beam intensity (for 20 s) as a function of an applied field frequency is shown in Fig. 9. The upper row shows the intensity integrated over the specular peak width $\alpha_f=0.388-0.455^\circ$. The bottom row shows the intensity at a specular peak maximum. The left and right columns correspond to spin up and down, respectively. One can see a resonance minimum centered at $f_{res}=26.2$~MHz with a FWHM of $\Delta f/f_{res}=7.7 \%$. The respective resonance deep (normalized on the counts outside the resonance) consists of for spin up: $18.7\pm2.7$~(\%) for the integrated count and $54.0\pm5.3$~(\%) for the maximum. For the spin down case, it is $16.4\pm4.4$~(\%) and $56.2\pm8.3$~(\%) for the maximum signal. One can see that the resonance dip for spin up and down coincide in resonance frequency within the error bar.

At the neutron magnetic spin resonance, the frequency of oscillating magnetic induction in the permalloy film (which is the frequency of the applied external magnetic field $f_{res}=26.2$~MHz is equal to the frequency of Larmor precession of neutron spin around the vector of magnetic induction with magnitude $B$: $f_L=\gamma B$ , where $\gamma=2 \mu /\hbar=2916 G^{-1} s^{-1}$ , $\mu$ is the magnetic moment of the neutron, and $\hbar = h/(2\pi)$ is the reduced Planck constant. The resonance frequency $f_{res} = 26.2$~MHz found in our experiments corresponds to Larmor precession in a magnetic induction of B = 9.0 kG. Note that the bias external field was about 20 Oe, therefore the sample exhibits a magnetic domain state. For this non-saturated magnetic state, the measured value of magnetic induction is close to the saturated magnetization 9.5 kG obtained by VSM and coincides with the magnetization 9.0 kG fitted from a reflectivity curve measured in a high magnetic field 4.4 kOe.  At contrast, for non saturated magnetic states the polarized neutron reflectometry method measures the projection of the magnetization on the direction of the applied field averaged over the ensemble of the domains. As a result, the reflectometry method provides a reduced magnetization of 4.0 kG in the non saturated magnetic state for the same external applied field of 20 Oe.  It is then clear, that the neutron spin resonance method derives the saturated magnetization 9.0 kG in the magnetic non saturated state, through the maximization of the spin-flip probability when the external oscillating field frequency matches to the Larmor precession.        

\begin{figure}[ht]
       \includegraphics[clip=true,keepaspectratio=true,width=1\linewidth]{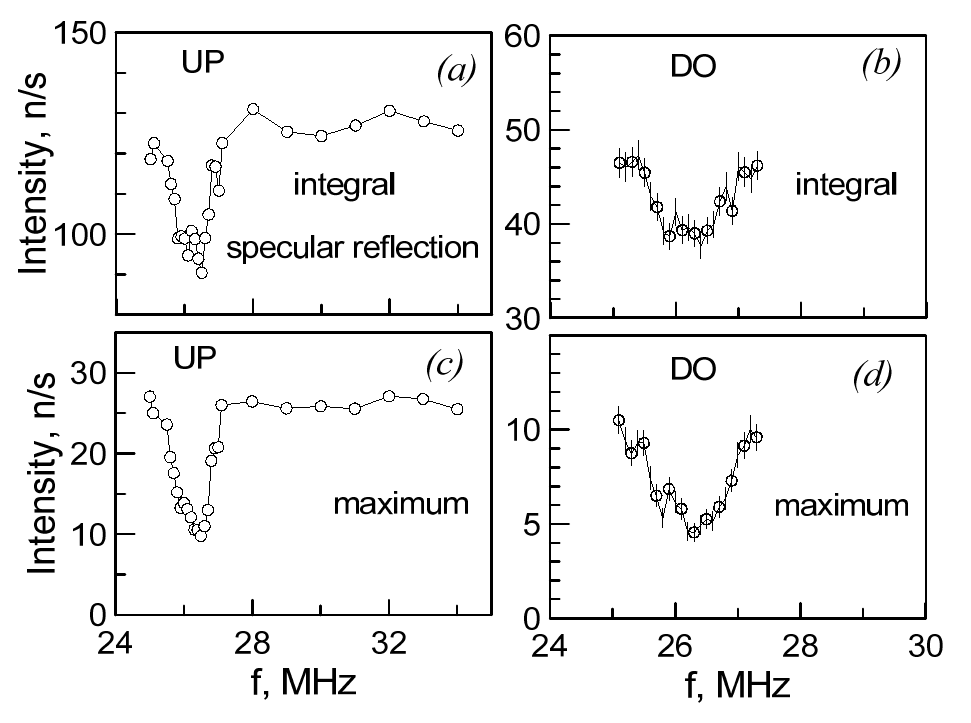}
          \caption{Specular reflection intensity as a function of the frequency of the oscillating applied magnetic field.}
   \label{fig9}
 \end{figure}

Thus, the method of reflectometry provides a measure of an averaged magnetization in a non-saturated state while the method of neutron magnetic spin resonance can measure the magnetic induction in individual domains even in a non-saturated state of the film. This possibility can be used for investigation of complex magnetic structures using neutron magnetic spin resonance. The accuracy of this method is defined by the width of the resonance (4.6 \% FWHM) which depends on the reflectometer resolution and also on the properties of the material.

\section{Conclusions}

We have presented three neutron scattering methods which allow the determination of the magnetic induction in absolute value in magnetic films: the Larmor precession, the Zeeman spatial beam-splitting and the neutron spin resonance. For the Larmor precession and the beam-splitting we have compared experiments performed on similar films of the same material FeAlSi. We have shown that these films are homogeneous and uniformly magnetized. The neutron methods have the advantages that they have the potential to provide information about the homogeneity of the magnetization of the film through its thickness. The refraction method in particular can easily identify differences in magnetization at the top or bottom interfaces of a system. Besides, the neutron beam-splitting method presented here is not sensitive to the thickness of the film and thus an accurate knowledge of the film thickness is not necessary in contrast to bulk magnetometry. Such techniques could find applications to the study of thick films of permanent magnets.
The neutron spin resonance was used for the investigation of a non-saturated magnetic film. The magnetic state of the film was characterized by polarized neutron reflectometry and VSM which are averaging methods over an ensemble of the magnetic domains. The method of neutron spin resonance provides the value of magnetic induction inside an individual magnetic domain. This is a unique way of obtaining such quantitative information on the induction value compared to other techniques such as Kerr microscopy or MFM.

 \acknowledgments

The authors are thankful to E.B. Dokukin and A.V. Petrenko for the help in Larmor precession experiment, J. Franke, J. Major and A. R\"uhm for the help in neutron spin resonance experiments. The authors acknowledge V.L Aksenov, V.K. Ignatovich, Yu.V. Nikitenko and T. Keller for fruitful discussions. This work has been supported by the Russian Fund of Basic Research project -12-02-00003-a.


\newpage

\subsection*{Appendix A. Refraction 'vacuum-film'}
\appendix
\numberwithin{equation}{section}
\begin{multline}\\
 \frac{\hbar p_i^2}{2m} +\mu B_0=  \frac{\hbar {(p^{+-}_f)}^2}{2m}-\mu B_1+V_1~(A.1)\\
{(p^{+-}_f)}^2=p^2_i- \frac{2m}{\hbar} [V_1-\mu(B_0+B_1)]~(A.2)\\
{(p^{--}_f)}^2=p^2_i- \frac{2m}{\hbar} [V_1+\mu(B_0-B_1)]~(A.3)\\
{(p^{++}_f)}^2=p^2_i- \frac{2m}{\hbar} [V_1-\mu(B_0-B_1)]~(A.4)\\
{(p^{-+}_f)}^2=p^2_i- \frac{2m}{\hbar} [V_1+\mu(B_0+B_1)]~(A.5)\\
{(\theta^{+-}_f)}^2=\theta^2_i- \frac{2m}{\hbar} [V_1-\mu(B_0+B_1)]\lambda^2~(A.6)\\
{(\theta^{++}_f)}^2=\theta^2_i- \frac{2m}{\hbar} [V_1-\mu(B_0-B_1)]\lambda^2~(A.7)\\
{(\theta^{--}_f)}^2=\theta^2_i- \frac{2m}{\hbar} [V_1+\mu(B_0-B_1)]\lambda^2~(A.8)\\
{(\theta^{-+}_f)}^2=\theta^2_i- \frac{2m}{\hbar} [V_1+\mu(B_0+B_1)]\lambda^2~(A.9)\\
  \notag
 \end{multline}
\\

\subsection*{ Appendix B.  Refraction 'film-substrate'}
\appendix
\numberwithin{equation}{section}
\begin{multline}\\
 \frac{\hbar p_i^2}{2m} +\mu B_2+V_2=  \frac{\hbar {(p^{+-}_f)}^2}{2m}-\mu B_1+U~(B.1)\\
{(p^{+-}_f)}^2=p^2_i+ \frac{2m}{\hbar} [(V_2-U)+\mu(B_2+B_0)]~(B.2)\\
{(p^{++}_f)}^2=p^2_i+ \frac{2m}{\hbar} [(V_2-U)+\mu(B_2-B_0)]~(B.3)\\
{(p^{--}_f)}^2=p^2_i+\frac{2m}{\hbar} [(V_2-U)-\mu(B_2-B_0)]~(B.4)\\
{(p^{-+}_f)}^2=p^2_i+\frac{2m}{\hbar} [(V_2-U)-\mu(B_2+B_0)]~(B.5)\\
{(\theta^{+-}_f)}^2=\theta^2_i+ \frac{2m}{\hbar} [(V_2-U)+\mu(B_2+B_0)]\lambda^2~(B.6)\\
{(\theta^{++}_f)}^2=\theta^2_i+ \frac{2m}{\hbar} [(V_2-U)+\mu(B_2-B_0)]\lambda^2~(B.7)\\
{(\theta^{--}_f)}^2=\theta^2_i+ \frac{2m}{\hbar} [(V_2-U)-\mu(B_2-B_0)]\lambda^2~(B.8)\\
{(\theta^{-+}_f)}^2=\theta^2_i+ \frac{2m}{\hbar} [(V_2-U)-\mu(B_2+B_0)]\lambda^2~(B.9)\\
  \notag
 \end{multline}

\end{document}